\documentstyle[aps,prl]{revtex}
\begin{document}
\input{psfig}
\draft
\title{\bf Nonextensive approach to decoherence in quantum mechanics}
\author{A. Vidiella-Barranco \footnote{Permanent address:Instituto de 
F\'\i sica ``Gleb Wataghin'',
Universidade Estadual de Campinas, 13083-970   Campinas  SP  Brazil}
\footnote{Email address: vidiella@ifi.unicamp.br} and 
H. Moya-Cessa\footnote{Email address: hmmc@inaoep.mx}}
\address{ INAOE, Coordinaci\'on de Optica, 
Apdo. Postal 51 y 216, 72000 Puebla, Pue., Mexico.}
\date{\today}
\twocolumn
\maketitle
\begin{abstract}
We propose a nonextensive generalization ($q$ parametrized) of the von Neumann 
equation for the density operator. Our model naturally leads to the phenomenon
of decoherence, and unitary evolution is recovered in the limit of 
$q\rightarrow 1$. The resulting evolution yields a nonexponential decay
for quantum coherences, fact that might be attributed to nonextensivity. We
discuss, as an example, the loss of coherence observed in trapped ions.
\end{abstract}
In the past decade, there have been substantial advances regarding a nonextensive,
$q$-parametrized generalization of Gibbs-Boltzmann statistical mechanics 
\cite{tsal1,tsal2}.
The $q$ parametrization is based on an approximate expression for the
exponential, or the $q$-exponential
\begin{equation}
e_q(x)=\left[1+(1-q)x\right]^{1/(1-q)},\label{expapp}
\end{equation}
being the ordinary exponential recovered in the limit of $q\rightarrow 1$
($e_1(x)=e^x$). Thus we have a nonextensive $q$-exponential, i.e., 
$e_q(x)e_q(y)\neq e_q(x+y)$, in general.
Such a parametrization is the basis of Tsallis statistics, in which
it is used a $q$-parametrized natural logarithm for the definition of
a generalized entropy. 
We recall that in Tsallis' statistics the entropy follows the {\it pseudo-additive rule}
\cite{tsal1,tsal2} $S_q(A+B)=S_q(A)\break +S_q(B)+(1-q)S_q(A)S_q(B)$, where $A$ and $B$ 
represent two independent systems, i.e., $p_{ij}(A+B)=p_i(A)p_j(B)$. Hence for $q<1$ we 
have the supper-additive regime, and for $q>1$ the sub-additive regime. The parameter
$q$ thus characterizes the {\it degree of extensivity} of the physical system considered.
Tsallis' entropy seems to be previligied, in the 
sense that it satisfies the concavity criterion, i.e., it has a well defined 
concavity \cite{tsal1,renio1}. Moreover, it has also been shown that most of the formal
structure of the standard statistical mechanics (and thermodynamics)
is retained within the nonextensive statistics, e.g., the H-theorem \cite{mariz}, and the 
Onsager reciprocity theorem \cite{cacer}.
Nonextensive effects are usual in many systems exhibiting, for instance, long-range
interactions, and the nonextensive formalism has been succesfully applied not only to 
several interesting physical problems, but also outside physics \cite{tsal2}. We may
cite for instance, applications to statistical mechanics \cite{renio2}, L\'evy anomalous
supperdiffusion \cite{tsal3}, quantum scattering \cite{ion}, low dimensional dissipative
systems \cite{tsal4}, in cosmology \cite{plast}, and others \cite{tsal2}. 
This means that 
departures from exponential (or logarithmic) behavior not so rarely occur in nature, and
that the parametrization given by Tsallis' statistics seems to be adequate for treating them.
Discussions concerning the quantification of quantum entanglement \cite{avb,rajago1} as well as 
the implications on local realism \cite{rajago2} may be also found, within the nonextensive 
formalism. Quantum entanglement has itself a nonlocal nature, and this may have somehow a 
connection with the general idea of nonextensivity. In fact, it has been concluded that 
entanglement may be enhanced in the supper-additive regime (for $q<1$) \cite{avb,rajago1}. 
Thus, although nonextensive ideas open up new interesting possibilities, their application
to a field such as foundations of quantum mechanics has not been so frequent. 
We would like, in this letter, to address fundamental aspects of quantum theory within a 
nonextensive approach.

In quantum mechanics, the issue of decoherence, or the transformation of pure states 
into statistical mixtures has been atractting a lot attention, recently,
especially due to the potential applications, such as quantum computing and quantum criptography 
\cite{comp}, that might arise in highly controlled purely quantum mechanical systems, e.g.,
in trapped ion systems \cite{wine}. That extraordinary control 
is also allowing to address fundamental problems in quantum mechanics, such as 
the Schr\"odinger cat problem \cite{found}, and the question of the origin of decoherence itself.
Despite of the progresses, it does not exist yet a proper theory handling the
question of the loss of coherence in quantum mechanics, although there are
several propositions, involving either some coupling with an
environment \cite{zurek,giuli}, or spontaneous (intrinsic) mechanisms \cite{giuli}. 
Indeed dissipative environments, in which it is verified loss of energy,
tend to cause decoherence. In spite of the destructive action of dissipation, though,
it has been worked out a scheme to recover quantum coherences in cavity fields \cite{ours}, 
even if its environment is at $T\neq 0$, and after the system's energy, 
and coherences, have substantially decayed. 
Regarding the intrinsic decoherence, several models have been already presented 
\cite{giuli}, and decoherence has been attributed, for instance, to stochastic jumps in the 
wave function \cite{milburn}, or gravitational effects \cite{ellis}. These models may contain 
one \cite{milburn} or even two new parameters \cite{ghirardi}. It is normally proposed some 
kind of modification of the von Neumann equation for the density operator $\hat{\rho}$
\begin{equation}
\frac{d\hat{\rho}}{d t}=-\frac{i}{\hbar}[\hat{H},\hat{\rho}].\label{vne}
\end{equation}

A typical model yielding intrinsic decoherence, such as 
Milburn's \cite{milburn} (see also reference \cite{bonifacio}), gives the following modified
equation for the evolution of $\hat{\rho}$ 
\begin{equation}
\frac{d\hat{\rho}}{d t}\approx-\frac{i}{\hbar}[\hat{H},\hat{\rho}]-
\frac{\tau}{2\hbar^2}[\hat{H},[\hat{H},\hat{\rho}]],\label{milbeq}
\end{equation}
where $\tau$ is a fundamental time step. The second term in the right-hand-side (double
commutator), leads to an exponential decay of coherences without energy loss \cite{milburn,moya}.
Nevertheless, other models concerning decoherence \cite{ellis}, as well as quantum 
measurements \cite{caves}, predict nonexponential decay of coherences, going as 
$\exp(-\gamma t^2)$ rather than as $\exp(-\gamma t)$. In \cite{ellis}, decoherence  
is attributed to a whormhole-matter coupling of nonlocal nature. This suggests 
that decoherence might not be appropriately described by a Markovian stochastic process
\cite{milburn}. Moreover, experimental evidence of nonexponential decay
of quantum coherences has been recently reported, in a experiments involving trapped 
ions, \cite{wine2}, as well as in quantum tunneling of cold atoms in an 
accelerating optical potential \cite{wilk}. 

Here we propose a novel model to treat decoherence, within a nonextensive realm. 
We are able to write down a $q$-parametrized (generalized) evolution for the density operator, 
which naturally leads to decoherence. As a first step we may express von Neumann's equation
[Eq. (\ref{vne})] in terms of the superoperator $\hat{\cal L}$, being 
$i\hbar\hat{\cal L}\hat{\rho}=\hat{H}\hat{\rho}-\hat{\rho}\hat{H}$,
so that we may write its formal solution
in a simple form
\begin{equation}
\hat{\rho}(t)=\exp(\hat{\cal L}t)\hat{\rho}(0).
\end{equation}
Now, we substitute the exponential superoperator $\exp(\hat{\cal L}t)$ by a
$q$-parametrized exponential as defined in Eq. (\ref{expapp})
\begin{equation}
\hat{\rho}(t)=\left[1+(1-q)\hat{\cal L}t\right]^{1/(1-q)}\hat{\rho}(0).\label{genevol}
\end{equation}
The standard unitary evolution given by von Neumann's equation is recovered in the 
limit of $q\rightarrow 1$. Hence, departing from Eq. (\ref{genevol}), we may write a 
generalized ($q$-parametrized) von Neumann equation, or
\begin{equation}
\frac{d\hat{\rho}}{d t}=\frac{\hat{\cal L}}{1+(1-q)\hat{\cal L}t}\,\hat{\rho}(t).
\label{genvne}
\end{equation}
Here we consider the case in which the parameter $q$ is very close to one. This 
represents a small deviation from the unitary evolution given by (\ref{vne}), but yields
loss of coherence. For $|1-q|\Omega t\ll 1$, where $\Omega$ is 
the characteristic frequency of the physical system considered, we may write
\begin{equation}
\frac{d\hat{\rho}}{d t}\approx \left[\hat{\cal L}-(1-q)t\hat{\cal L}^2\right]\,
\hat{\rho}(t).\label{approxeq0}
\end{equation}
It may be shown, from Eq. (\ref{approxeq0}), that there is decay of the nondiagonal elements 
in the basis formed by the eigenstates of $\hat{H}$, characterizing decoherence, while the 
diagonal elements are not modified. Therefore the density operator remains normalized at all 
times, or $Tr\hat{\rho}(t)=Tr\hat{\rho}(0)=1$, property which must hold in any proper model 
for decoherence. We shall remark that in order to have a physically acceptable dynamics,
the extensivity parameter should be greater than one ($q>1$), i.e., in the sub-additive regime.
We may also rewrite Eq. (\ref{approxeq0}) as
\begin{equation}
\frac{d\hat{\rho}}{dt}\approx-\frac{i}{\hbar}[\hat{H},\hat{\rho}]+g(t)
[\hat{H},[\hat{H},\hat{\rho}]],
\label{approxeq}
\end{equation}
where $g(t)=(1-q)t/\hbar^2$. Eq. (\ref {approxeq}) is
similar to Eq. (\ref{milbeq}), apart from the time-dependent
factor $g(t)$ multiplying the double commutator. Note that Eq. (\ref{approxeq}) is only valid
for short times, or $|1-q|\Omega t\ll 1$. We have therefore constructed a novel model, 
based on a nonextensive generalization of von Neumann's equation, which predicts loss of 
coherence, although with a time-dependence for the nondiagonal elements in the energy basis 
different from most that can be found in the literature \cite{giuli}. 
This particular time-dependence will lead to a {\it nonexponential decay of quantum coherences}, 
as we are going to discuss.

Now we would like to apply our model to a specific system in which experimental results are 
available. One of the most interesting systems investigated nowadays is a single trapped ion 
interacting with laser fields. Quantum state engineering of motional states of a massive 
oscillator has been achieved \cite{wine1}, and the loss of coherence in that system has 
been already observed \cite{wine1,wine2}. In the experiment described in reference \cite{wine1}, 
two Raman lasers induce transitions between two internal electronic states ($|e\rangle$ and 
$|g\rangle$)
of a single $^9$Be$^+$ trapped ion, as well as amongst center-of-mass vibrational states 
$|n\rangle$. In the interaction picture, and under the rotating wave approximation, 
the effective hamiltonian (in one dimension) may be written as \cite{wine1}
\begin{equation}
\hat{H}^I_{int}=\hbar \Omega\left( \sigma_+ e^{i\eta(\hat{a}^\dagger_x+\hat{a}_x)-i\delta t}+
\sigma_- e^{-i\eta(\hat{a}^\dagger_x+\hat{a}_x)+i\delta t}\right),\label{hamilt}
\end{equation}
being $\Omega$ the coupling constant, $\delta$ the detuning of the frequency difference of the
two laser beams with respect to $\omega_0$, which is the energy difference between 
$|e\rangle$, or $|g\rangle$ ($E_e-E_g=\hbar\omega_0$). The Lamb-Dicke parameter 
is $\eta=k\sqrt{\hbar/(2m\omega_x)}$, where $k$ is the magnitude of the difference between the
wavevectors of the two laser beams, $\omega_x$ is the vibrational frequency of the ion's center of
mass having a mass $m$, and $\hat{a}_x(\hat{a}^\dagger_x)$, are annihilation and creation 
operators of vibrational excitations. The ion is Raman cooled to
the (vibrational) vacuum state $|0\rangle$, and from that Fock (as well as coherent and squeezed) 
states may be prepared 
by applying a sequence of laser pulses \cite{wine1}. If the atom is initially in the $|g,n\rangle$ 
state, and the Raman lasers are tuned to the first blue sideband ($\delta=\omega_x$), the 
evolution according to the hamiltonian in Eq. (\ref{hamilt}) (for small $\eta$) will be such
that the system will perform Rabi oscillations between the states $|g,n\rangle$ and
$|e,n+1\rangle$. If the excitation distribution of the initial vibrational state of the ion
is $P_n$, occupation probability of the ground state $P_g(t)$ will be given by
\begin{equation}
P_g(n,t)=\frac{1}{2}\left[1+\sum_n P_n \cos(2\Omega_nt)\,e^{-\gamma_nt}\right].\label{pnwine}
\end{equation}
Here the Rabi frequency $\Omega_n$ is
\begin{equation}
\Omega_n=\Omega\frac{e^{-\eta^2/2}}{\sqrt{n+1}}\eta L_n^1(\eta^2),
\end{equation}
where $\Omega/2\pi=500$kHz, $L_n^1$ are Laguerre generalized polynomials, 
and $\gamma_n=\gamma_0(n+1)^{0.7}$, with $\gamma_0=11.9$kHz. 

The experimental results are fitted using Eq. (\ref{pnwine}) \cite{wine1}. 
The Rabi oscillations are damped,
and an empirical damping factor, $\gamma_n$, is introduced in order to fit the 
experimental data \cite{wine1}. There heve been attempts to derive such an unusual dependence on $n$
for the damping factor $\gamma_n$, by taking into account the fluctuations in laser fields and the 
trap parameters \cite{knight}. Those models are somehow connected to a evolution of the type given by
Eq. (\ref{milbeq}) \cite{milburn,moya}, which predicts loss of coherence without loss of 
energy. In fact, at a time-scale in which the ion's energy remains almost constant,  
decoherence is considerably large \cite{wine,wine1}. Moreover, the actual causes of loss of
coherence in experiments with trapped ions are still not identified \cite{wine,wine3}.
Our approach differs from previous ones because we depart from a different dynamics which
leads to a peculiar nonexponential time-dependence. By employing a similar methodology to
the one discussed in reference \cite{moya}, we may calculate, from Eq. (\ref{approxeq}), the 
probability of the atom to occupy the ground state, obtaining
\begin{equation}
P_g^q(n,t)=\frac{1}{2}\left[1+\sum_n \cos(2\Omega_nt)\,e^{-\gamma_{n,q}t^2}\right],
\label{pnours}
\end{equation}
where $\gamma_{n,q}=(q-1)\Omega_n^2/2$, i.e., the damping factor arising in our model
depends on the Rabi frequency $\Omega_n$ and on the parameter $q$ in a simple way.
It is also verified a nonexponential decay which goes as $\exp(-\gamma t^2)$ 
rather than an exponential one. Now we may proceed with a graphical comparison between
the curves plotted from expressions (\ref{pnwine}) and (\ref{pnours}). This is shown in 
Figure 1, for two distinct cases. In Figure 1a we have 
plotted both the probability $P_g$ in Eq. (\ref{pnours}) (dashed line) and the one in
Eq. (\ref{pnwine}) (full line), as a function of time, for an initial state $|g,0\rangle$ 
$(P_n=\delta_{n,0})$, a Lamb-Dicke parameter $\eta=0.202$, and $q=1.001$. We note a clear departure
from exponential behavior in the curve arising from our model, although they might coincide for some
time-intervals. We remark that Eq. (\ref{pnours}) is valid for times such that $|1-q|\Omega t\ll 1$.
Here $|1-q|\Omega t_{max}\approx 0.17$. Right below, in Figure 1b, there is a similar plot, but 
using a different initial condition for the distribution of excitations of the ion state, e.g., a 
coherent state $|\alpha\rangle$, with $\overline{n}=\alpha^2=3$
($P_n=\exp(-\alpha^2)\alpha^{2n}/n!$) instead of the vacuum state 
$|0\rangle$. In the coherent state case both curves are very close, although it seems that our
curve (dashed line) represents a better fit for the experimental results \cite{wine1}. 
In general, our results are in reasonable agreement with the experimental data, 
as it may be seen in Figure 1 and in reference \cite{wine1}. This means that a nonexponential 
decay may also account for the loss of coherence in the trapped ions system.

\begin{figure}[hp]
\vspace{0.1cm}
\centerline{\hspace{1.0cm}\psfig{figure=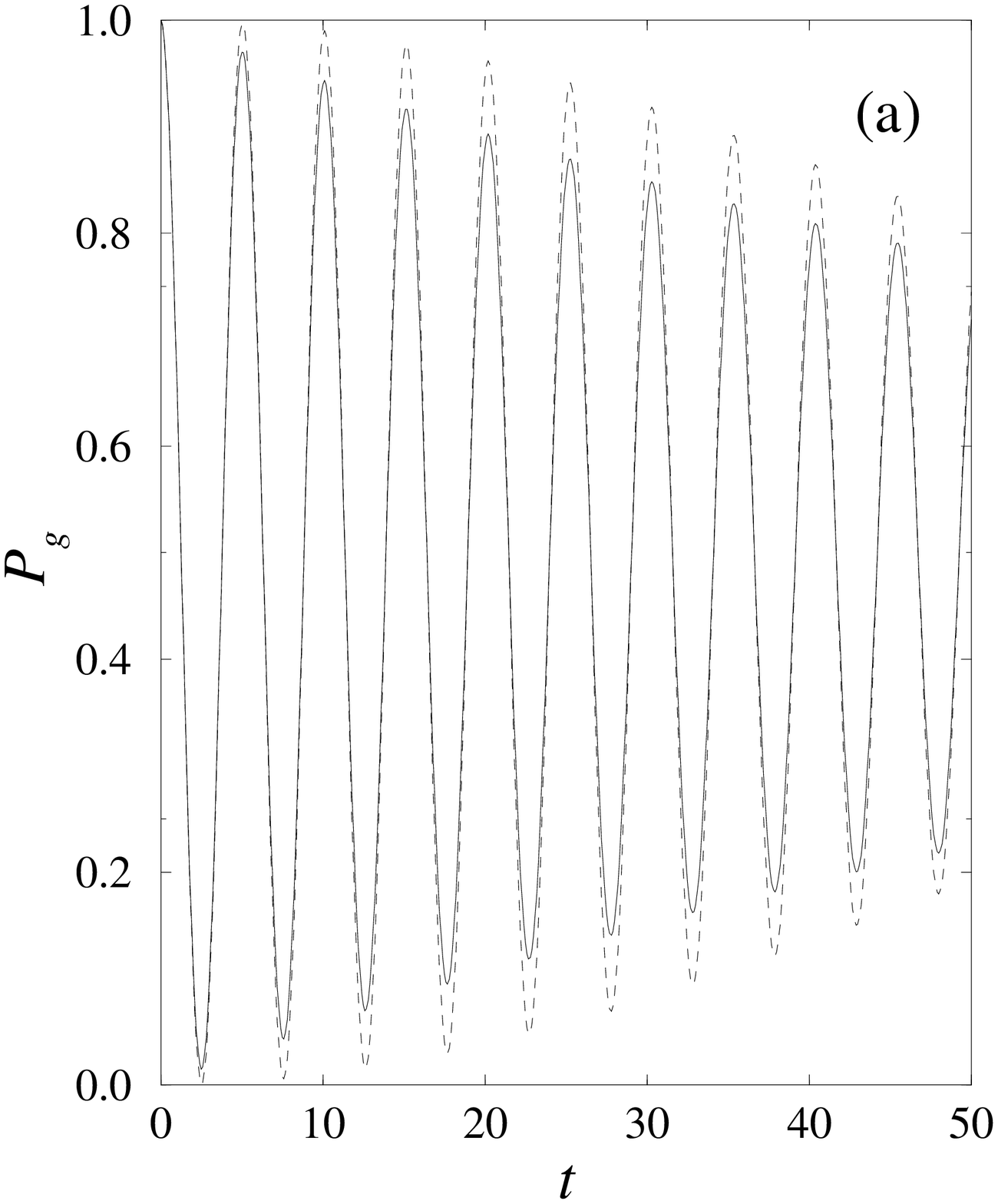,height=4cm,width=6cm}}
\end{figure}
\vspace{0.2cm}
\begin{figure}[hp]
\vspace{0.1cm}
\centerline{\hspace{1.0cm}\psfig{figure=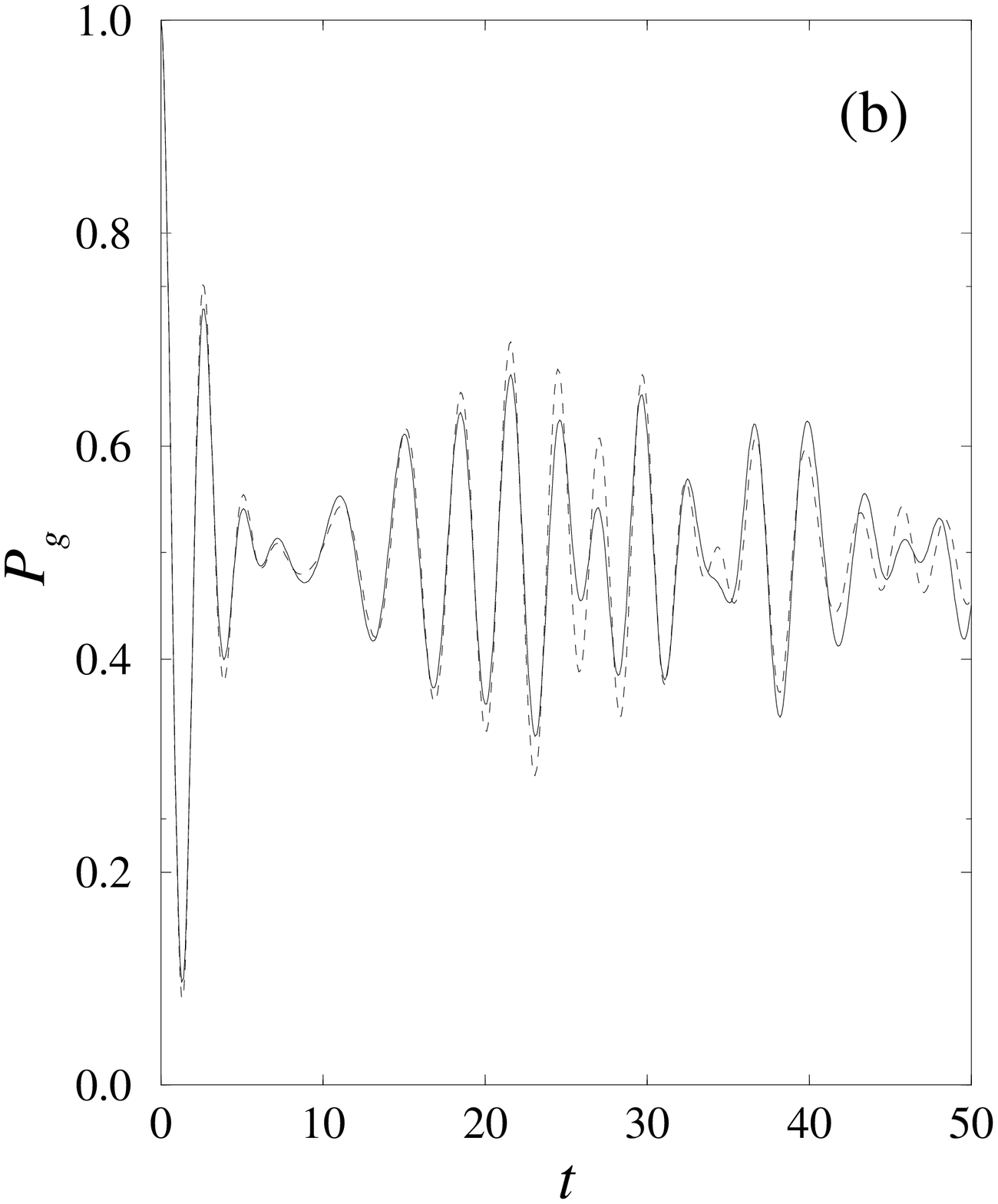,height=4cm,width=6cm}}
\vspace{0.7cm}
\caption{Probability of having the atom in the ground state $P_g$ as a 
function of time;(a) for an initial (vibrational) vacuum state and (b) for 
an initial (vibrational) coherent state with $\overline{n}=3$.
In either case $\eta=0.202$, and $\Omega/2\pi=500$kHz. 
The solid lines are calculated from from Eq.(10), and the 
dashed lines from Eq.(12), with an extensivity parameter $q=1.001$.}
\end{figure}

In our model, the effects leading to decoherence are embodied within the nonextensive 
parameter $q$, rather than into a ``fundamental time step'' $\tau$, as it 
occurs in Milburn's model. Nonextensivity may be especially relevant when 
nonlocal effects, for instance, long-range interactions, or memory effects are involved 
\cite{tsal2}. In the specific example treated here, the ion is not completely isolated 
from its surroundings; electric fields generated in the trap electrodes 
couple to the ion's charge, and this is considered to be a genuine source of decoherence 
to the vibrational motion of the ion \cite{wine,wine3}.

In conclusion, we have presented a novel approach for treating the loss of coherence in quantum
mechanics, based on a nonextensive formalism. In our model, the quality of decoherence 
depends on a single parameter, $q$, related to the nonextensive properties of the physical 
system considered. We obtain, with such a parametrization, an evolution equation which is a
(nonextensive) generalization of von Neumann's equation for the density operator, and that 
leads, in general, to a nonexponential decay of quantum coherences. We have
applied our model to a concrete physical problem, that is, the decoherence occurring in the
quantum state of a single trapped ion undergoing harmonic oscillations. This is a well know case
in which decoherence may be readily tracked down, yet its causes remain unexplained. We have found
that for values of the parameter $q$ rather close to one (in the sub-additive regime), our model is 
in reasonable agreement with the available experimental results, without the need of introducing any 
supplementary assumptions.

\acknowledgements

One of us, H.M.-C., thanks R. Bonifacio and P. Tombesi for useful comments.
A.V.-B. would like to thank the hospitality at INAOE, M\'exico.
This work was partially supported by CONACyT (Consejo Nacional de
Ciencia y Tecnolog\'\i a, M\'exico), and CNPq (Conselho Nacional de 
Desenvolvimento Cient\'\i fico e Tecnol\'ogico, Brazil).

% \begin{figure}
% \caption{}
% \label{}
% \end{figure}

% \begin{table}
% \caption{}
% \label{}
% \end{tabular}
% \end{table}

\end{document}